# Cooling rate effect on thermoremanent magnetization in archaeological baked clays: an experimental study on modern bricks

Gwenaël Hervé[1], Annick Chauvin[2], Philippe Lanos[2,3], Pierre Rochette[1], Mireille Perrin[1], Michel Perron d'Arc[4]

[1] Aix Marseille Univ, CNRS, IRD, INRA, Coll France, CEREGE, Aix-en-Provence, France

[2] Univ Rennes, CNRS, Géosciences Rennes, UMR 6118, 35000 Rennes, France

[3] IRAMAT-CRP2A, Univ Bordeaux Montaigne, CNRS, Bordeaux, France

[4] Amphoralis, Musée des potiers gallo-romains, Sallèles-d'Aude, France

Corresponding author: gwenael.herve@u-bordeaux-montaigne.fr

Now at: IRAMAT-CRP2A, LabEx Sciences archéologiques de Bordeaux (LaScArBx), Univ Bordeaux Montaigne, CNRS, Bordeaux, France and Univ Rennes, CNRS, Géosciences Rennes, UMR 6118, 35000 Rennes, France

Abstract

The influence of cooling rate on the intensity of thermoremanent magnetization (TRM) and the necessity to correct archaeo/palaeointensities for this effect have long been recognized. However the reliability of the correction is still questioned. We studied 35 bricks baked in two modern kilns (SK and BK) in known experimental conditions and with measurements of the direction and intensity of the geomagnetic field at the site. The smallest kiln (SK, 0.2 m$^3$) cooled in around 12 hours and the biggest kiln (BK, 8 m$^3$) in around 40 hours. Thermomagnetic, hysteresis and backfield curves indicated that the main magnetic carriers were Ti-poor titanomagnetites and Ti-poor titanohematites. The fraction of the TRM carried by Ti-poor titanohematites is the main difference between the two sets of bricks. This fraction is around 5-10% in bricks from BK kiln and up to 40% in those from SK kiln. Intensities of the Earth's magnetic field were determined using the original Thellier-Thellier protocol with correction of TRM anisotropy. The average intensities overestimate the expected field intensity by 5% (SK) and 6% (BK). This result emphasizes the necessity of the cooling rate correction. In order to have a detailed evaluation of the cooling rate effect, we used several slow cooling rates: 0.8, 0.4, 0.2 and 0.1°C/min. The correction factors obtained with the 0.8°C/min cooling ranged between -2% and 21% and were proportional to the TRM fraction carried by Ti-poor titanohematite. The higher proportion of these grains in bricks from SK kiln led to an overestimate of the correction factor and an underestimate of the intensity by 7%. However, the expected intensity is recovered when temperature steps higher than 580°C (i.e. in the range of Ti-poor titanohematites unblocking temperatures) were excluded from the calculation of archaeointensity and cooling rate correction. In the case of the BK kiln bricks, for which Ti-poor titanohematites does not contribute significantly to the TRM, all tested cooling rates give average intensities close to the expected value. Incorrectly estimating the duration of the archaeological cooling has therefore a low impact on the accuracy of the archaeointensity data on these kinds of material.

1. Introduction

Beyond the first direct measurements in 1832, the secular variation of the absolute intensity of the geomagnetic field is recovered from archaeomagnetic and volcanic materials. During their last cooling from high-temperature, they acquired a thermoremanent magnetization (TRM) whose intensity is proportional to the ambient geomagnetic field strength. The most used palaeointensity protocols, the Thellier and Thellier (1959) and derived methods, are based on the comparison of the intensity of the natural remanent magnetization (NRM) with the one of a TRM acquired in a known laboratory field. Samples must obey the Thellier laws of reciprocity, independence and additivity of partial thermoremanent magnetization (pTRM) acquired in non-overlapping temperature intervals. Experiments are challenging, because several effects, such a coarse or unstable ferromagnetic mineralogy, can lead to a departure away from the Thellier laws (e.g. Dunlop, 2011). Moreover, the palaeointensity depends, especially in the case of archaeological baked clays, on the magnetic anisotropy (e.g. Veitch et al., 1984) and on the cooling duration (e.g. Fox and Aitken, 1980). Regarding the latter, a slower cooling gives more time to magnetic grains to reach equilibrium of magnetization, which slightly lowers the blocking temperature and therefore modifies the TRM intensity (e.g. Néel, 1955; Dodson and McClelland, 1980).

In order to accurately estimate the archaeointensity, the past cooling conditions have to be replicated throughout the Thellier-Thellier protocol. But laboratory furnaces usually cool in 0.5-2 hours with pulsed air, much faster than the initial archaeological cooling. The difference between the cooling durations often results in underestimate of the laboratory TRM intensity, i.e. overestimate of the archaeointensity between 0 and 10% (e.g. Genevey et al., 2008) and sometimes up to 20-25% (Hervé et al., 2017; Poletti et al., 2013). This inaccuracy emphasized the need of a correction for the cooling rate effect, in order to avoid biases in regional or global geomagnetic models (e.g. Poletti et al., 2018; Hervé et al., accepted).

The most used correction of Gomez-Paccard et al. (2006) consists in an additional heating step with a slower cooling whose duration is assumed to be close to the archaeological one. The main pitfall of this method is the estimation of the duration of the past cooling. Unlike volcanic glasses with relaxation geospeedometry (Leonhardt et al., 2006; Ferk et al., 2010), there is no independent physical measurement of the past rate for archaeological baked clays. Experimental archaeology provides information (Genevey and Gallet, 2002; Genevey et al., 2016; Morales et al., 2011) but the cooling duration intimately depends on the kiln morphology (e.g. size, shape, number and size of opening in walls), the firecraft techniques, and even meteorological conditions (Genevey et al., 2016). For a given archaeological feature, the variability of these factors makes approximative the analogy with the few experimental studies and one can wonder about the impact of a miscorrection on the average archaeointensity.

When the cooling rate was not tested, it is crucial to identify archaeointensity data with a low cooling rate effect because they are theoretically the most accurate. Rock magnetism techniques can be useful tools because the cooling rate effect is sensitive to the ferromagnetic mineralogy especially to the grain size (Papusoi et al., 1972a). According to theoretical studies (Papusoi, 1972a and b; Dodson and McClelland-Brown, 1980; Halgedahl et al., 1980; Berndt et al., 2017) and experimental analysis on synthetic samples (Fox and Aitken, 1980; McClelland-Brown, 1984; Biquand, 1994; Yu, 2011; Biggin et al., 2013), the TRM intensity of magnetite and hematite increases with the cooling duration for single domain (SD) grains, is stable for pseudo single domain (PSD) and slightly decreases for multidomain (MD) grains. The relationship of the cooling rate effect with the ferromagnetic mineralogy is likely more complex in archaeological baked clays, which are assemblages of different iron oxides of various grain sizes and have a wide spectrum of unblocking temperatures. To our knowledge, whether the cooling rate effect could depend on the type of the ferromagnetic phases has not yet been experimentally studied in archaeological baked clays.

The present archaeomagnetic study on bricks baked in two experimental kilns that cooled over a known duration has two objectives. We first investigated the relation between the cooling rate effect and the chemical composition of magnetic carriers. Secondly, we studied the impact on archaeointensity determinations of a wrong estimation of the duration of the archaeological cooling. After presenting the experimental kilns and sampling conditions, we described the laboratory methods, rock magnetic properties and archaeointensity results of the studied bricks. We finally discussed the influence of the ferromagnetic mineralogy on the cooling rate corrections and on the recovery of the expected geomagnetic field strength.

2. Presentation of experimental bakings

2.1 Description of experimental kilns

The two experimental kilns are located in Amphoralis Museum in Sallèles-d'Aude (Southern France, Lat: 43.274130°N, Long: 2.936934°E). They are built using Roman techniques with two overlayed chambers separated by a perforated plate (Fig. 1). The lower chamber is the combustion chamber, where is located the fireplace, and the upper is the baking chamber containing the baked clays. The two kilns differ by size and shape. The baking chamber of the smallest (~0.2 m$^3$) kiln (labelled SK) is almost cylindrical with a height of 0.90 m and a diameter of 0.75 m at the plate level and of 0.40 m at the top (Fig. 1a). The biggest kiln (labelled BK) has a volume of ~8 m$^3$ with a 2.0 x 2.0 m plate and a maximal height of 1.6 m (Fig. 1b). The door of the baking chamber of BK kiln was closed during experimental baking, whereas the top of SK kiln remained open.

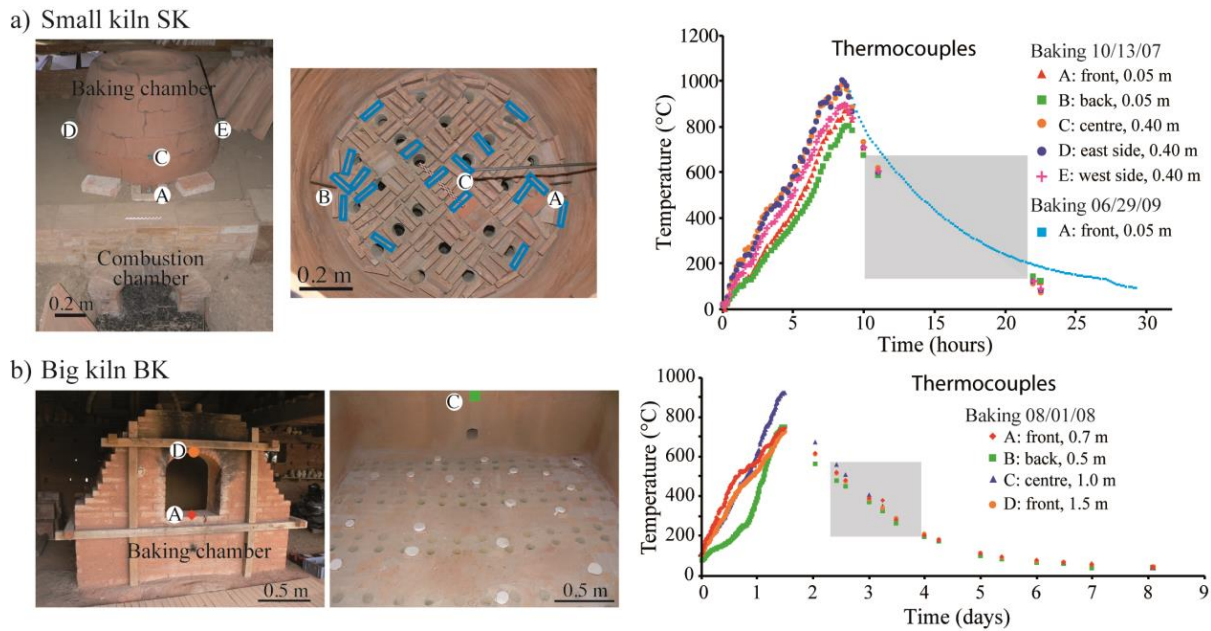

Figure 1: Photographs and heating-cooling temperature curves of the two experimental kilns a) SK small kiln and b) BK big kiln. The 17 sampled SK bricks are surrounded in light blue. The BK bricks were oriented *in situ* using the plaster cap method. The distance in meters associated to each thermocouple is the height above the plate. On the curves, the grey area indicates the time needed for the temperature to decrease in the range of unblocking temperatures of bricks.

Mixed archaeological objects, mainly bricks, tiles and potteries, were baked on 10/13/2007 in SK kiln and on 08/01/2008 in BK kiln. The geomagnetic field components were measured at the site and inside each kiln with a MEDA fluxgate magnetometer (FVM-400 handheld vector model).

The fluxgate was levelled horizontal with a bubble and, for outside measurements, azimuthally oriented using a sun compass. Measurements at the site and inside kilns were very consistent with the values predicted by IGRF12 model (Thébault et al., 2015), which highlighted the absence of a local geomagnetic anomaly (Tab. 1). The expected archaeointensity is 46.0 µT.

Table 1: Declination, inclination and intensity of the geomagnetic field predicted by the International Geomagnetic Reference Field model (IGRF12, Thébault et al., 2015) and measured at the site and inside each kiln using fluxgate magnetometer. Orienting the fluxgate in the horizontal plane was not possible inside the kilns.

|  | Dec. (°) | Inc. (°) | Int. (µT) |
| --- | --- | --- | --- |
| IGRF (10/13/2007) | -0.3 | 58.9 | 46.0 |
| IGRF (08/01/2008) | -0.2 | 58.9 | 46.0 |
| Measurement at the site | 0.8 | 58.8 | 46.1 |
| Inside small kiln (SK) | - | 59.0 | 46.0 |
| Inside big kiln (BK) | - | 60.0 ± 0.5 | 46.2 ± 0.4 |

The variation of the temperature during baking and cooling in SK and BK kilns were measured with five and four thermocouples respectively (Fig. 1). In the case of SK, the 10/13/2007 experiment could not have a continuous record of the temperature during cooling and information was completed with a posterior baking (06/29/09) (Fig. 1a). The temperature increased inhomogeneously between the different parts of the kilns and reached higher values in the centre of the baking chambers up to ~920°C in SK and ~1000°C in BK. The temperature exceeded 720°C in all thermocouples. Spatial variation of the temperature inside the kilns was lower during cooling, especially below the Néel and Curie temperatures of hematite and magnetite respectively when the baked clays acquired their TRM. This led us to assume a single cooling rate for each kiln. Cooling from 700°C to 100°C followed an exponential trend and took about 12 hours in the small SK kiln and about 60 hours in the big BK kiln.

2.2 Sampling

Sampling focused on small bricks baked on the plates of the two kilns. All have the same dimensions (80x55x18 millimeters) and were shaped from the same clays. In SK kiln, the 17 bricks were not oriented but their position on the side made possible the recovery of the inclination (Lanos, 1994). The 18 bricks from BK kiln were baked in the flat position (Fig. 1b). They were oriented *in situ* with a bubble level and a magnetic compass using the common archaeomagnetic plaster cap method. At the laboratory, bricks were cored in 1-inch diameter specimens.

3. Protocol

3.1 Rock magnetism

All measurements were performed at Géosciences Rennes palaeomagnetic laboratory, except hysteresis curves made at CEREGE laboratory. Remanent magnetizations were measured using a superconducting rock magnetometer (2G Enterprise) and a Molspin "Minispin" magnetometer. Low-field susceptibilities were measured with a Bartington MS2 susceptibility meter and thermomagnetic curves were performed, up to 680°C, with a KLY3-CS3 kappabridge on 13 bricks from each kiln.

Hysteresis curves up to 1 Tesla were performed at room temperature on chips from each brick using Princeton Micromag Vibrating Sampling Magnetometer. The variation of hysteresis curves with temperature was measured on 9 bricks with 22 steps from 100 to 660°C using the same device. The hysteresis curve at room temperature was measured a second time after heating, in order to check the absence of mineralogical alteration. Finally, backfield curves from 2.5 T of the isothermal remanent magnetization (IRM) were measured on 19 one-inch specimens using an ASC impulse magnetizer. These curves provided the $S_{300}$ ratio used as a proxy of the relative proportion of low and high coercivity minerals. This parameter was defined according to Stober and Thompson (1979) by:

$$S_{300} = - \frac{IRM\ at\ 0.3\ T\ after\ "saturation"\ at -2.5\ T}{IRM\ at -2.5\ T} \qquad (1)$$

3.2 Archaeointensity protocol

Intensities were determined on all bricks using the classical Thellier-Thellier method (Thellier and Thellier, 1959) with partial thermoremanent (pTRM) checks (Coe et al., 1978). Heatings were performed in a homemade furnace using 14 steps up to 650°C for SK bricks and using 11 steps up to 555°C for BK bricks.

At each temperature step, specimens were heated and cooled twice with the laboratory field $B_{lab}$ applied along their +z axis and then in the opposite sense, the z-axis being parallel to the thickness of the brick. The $B_{lab}$ value was fixed to 60 µT, which is approximately the average geomagnetic field strength during the last thousand years in Western Europe, rather than close to the expected present intensity. This choice better met the experimental conditions used for an archaeological feature cooled in unknown geomagnetic field strength. The effect of the anisotropy was corrected through the determination of the TRM anisotropy (ATRM) tensor (Veitch et al., 1984). We used the protocol of Chauvin et al. (2000) at 530°C with successive heatings in six orthogonal positions (+x, -x, +y, -y, +z, -z) followed by a stability check.

3.3 Cooling rate correction

At each step, cooling of the samples to room temperature lasted approximately 1.5 hours, much shorter than the initial cooling of the bricks in SK and BK kilns. The cooling rate effect on TRM intensity was estimated with four heating-cooling steps at the same temperature (Chauvin et al., 2000; Gomez-Paccard et al., 2006): (1) 1.5 hours cooling with + $B_{lab}$; (2) 1.5 hours cooling with - $B_{lab}$; (3) slow cooling with + $B_{lab}$; (4) same as (2) to monitor alteration. The TRMs acquired after each step are labelled $TRM_{q+}$, $TRM_{q1-}$, $TRM_{s+}$ and $TRM_{q2-}$. The cooling rate correction factor %corr is equal to:

$$\%corr = 100 \times \frac{TRM_{s+} - TRM_{q+}}{TRM_{q+}}$$

(2)

and the alteration factor %alt to:

$$\%alt = 100 \times \frac{TRM_{q2-} - TRM_{q1-}}{TRM_{q1-}}$$

(3)

The protocol was performed at 650°C for SK bricks and at 555°C for BK bricks, when around 85-90% of the NRM was demagnetized. Several slow cooling rates were successively tested. In the case of SK bricks, we used 0.8°C/min and 0.4°C/min slow cooling rates, which corresponded to cooling durations from 650°C to 100°C around 12 and 23 hours respectively. In the case of BK bricks, the slow cooling rates were set at 0.8°C/min, 0.4°C/min, 0.2°C/min and 0.1°C/min, which corresponded to cooling durations around 10, 19, 38 and 76 hours respectively.

Finally, after a further heating step at 650 or 555°C in zero field to erase the acquired pTRMs, additional 0.8°C/min slow coolings were performed from lower temperature steps: 385°C, 500°C and 580°C for SK bricks and 350°C, 430°C and 500°C for BK bricks. These additional experiments aimed to investigate variations of the cooling rate effect with temperature. All successive heating/cooling cycles are summarized in Fig. S1 in Supplementary Material.

## 4. Results

### 4.1 Rock magnetism

Even if they were made from the same clays, SK and BK bricks presented different rock magnetic properties (Fig. 2, Tab. S1). Thermomagnetic curves of BK bricks showed a single ferromagnetic phase with a pronounced Hopkinson peak (Fig. 2a). For most samples (11/13, e.g. BK-7 and BK-17), this phase had a Curie temperature between 550 and 570°C and was interpreted as a Ti-poor titanomagnetite. The Curie temperature of the two other samples (e.g. BK-30) was higher around 610°C.

The shapes of the thermomagnetic curves were less homogeneous in the case of SK bricks (Fig. 2b). Only three samples (e.g. SK-28) behaved in the same way as BK bricks with a sharp decrease of the susceptibility at 540-560°C. For six other samples (e.g. SK-20), the susceptibility decreased over a broader temperature range from 150-200°C. Finally, the curves of four samples (e.g. SK-4) highlighted a phase with Curie temperatures between 640 and 670°C that is likely a Ti-poor titanohematite.

Hysteresis and backfield curves of SK bricks (all except one) confirmed the mixing of the low coercivity Ti-poor titanomagnetite and of the high coercivity Ti-poor titanohematite (Fig. 2c). This phase was likely also present in most BK bricks, as seven (over nine) backfield curves did not saturate at low fields (Fig. 2d). The two other BK bricks had almost solely a low coercivity magnetic carrier, Ti-poor titanomagnetite in the case of BK-3 and partially oxidized magnetite or titanomaghemite in the case of BK-30, for which the Curie temperature was around 610°C.

In this study, the $S_{300}$ ratio appeared the most suitable way to quantify the fraction of Ti-poor titanohematite in bricks. With an average $S_{300}$ equal to -0.06 and 0.75 respectively, Ti-poor titanohematite was clearly more abundant in SK bricks than in BK bricks. According to the spectra of unblocking temperatures inferred from the Thellier-Thellier experiments, grains with unblocking temperatures higher than 600°C, i.e. Ti-poor titanohematite grains, carried up to 40% of the NRM in SK bricks (Fig. 2e). In most BK bricks, around 90% of the NRM was demagnetized at 555°C indicating that Ti-poor titanohematite only carried a few per cent of the NRM (Fig. 2f).

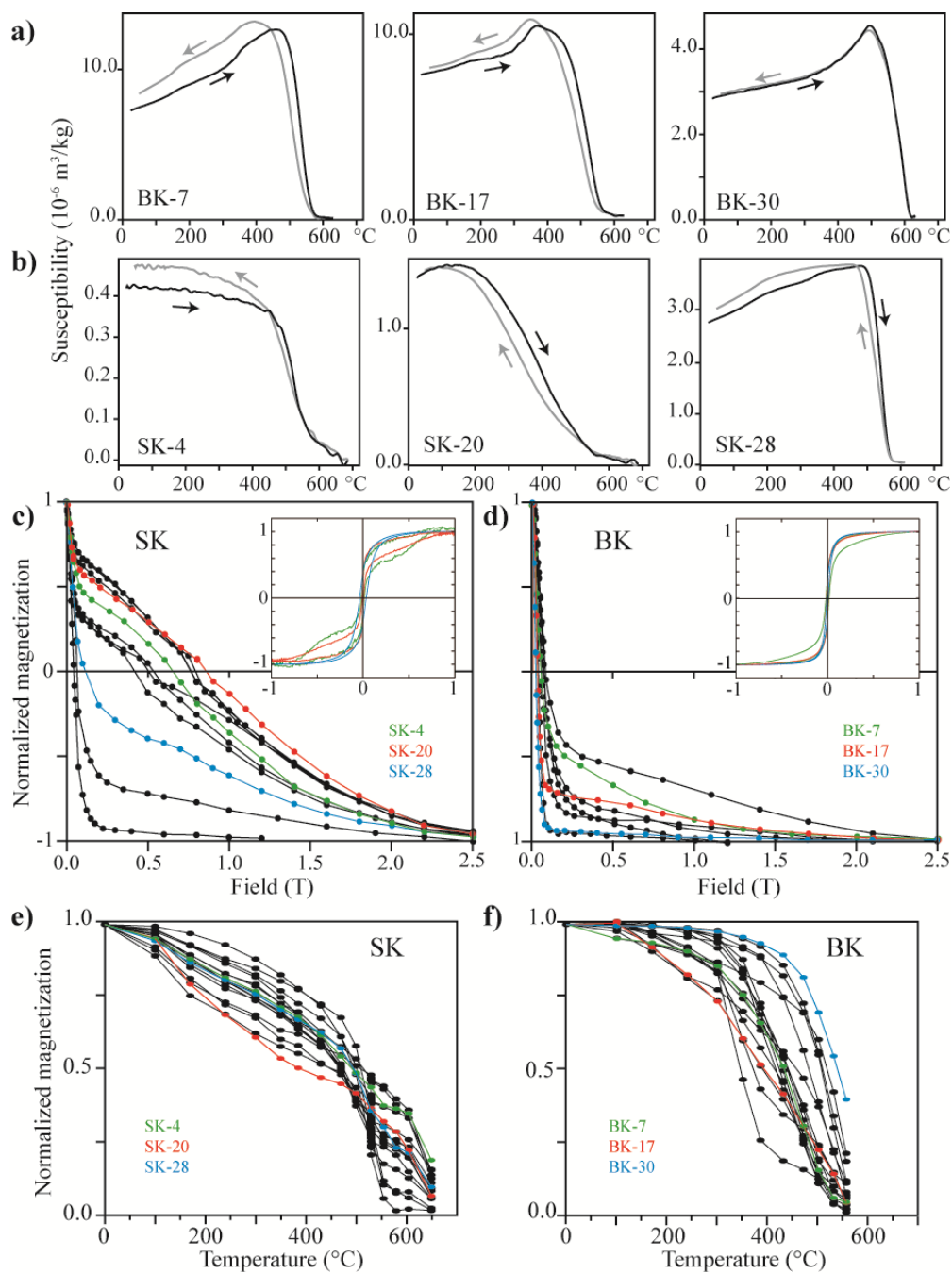

Figure 2: Rock magnetic properties of SK (b-c-e) and BK bricks (a-d-f) with representative thermomagnetic (a-b), IRM backfield and hysteresis (c-d) curves, and unblocking temperature spectra inferred from Thellier-Thellier experiments (e-f).

High-temperature measurements of hysteresis curves precised the impact of high coercivity minerals on the remanent magnetization. For two of the nine measured bricks, an increase of the saturation remanent magnetization ($M_{rs}$) from 300°C revealed mineralogical alteration during heating. The good agreement between hysteresis curves performed at room temperature before and after heating excluded alteration in the seven other bricks. We observed two types of variations of the remanence ratio ($M_{rs}/M_s$) and coercive force ($B_c$) with temperature (Figure 3). Three bricks (SK-28, BK-7 and BK-30) showed a decrease tending to zero of $M_{rs}/M_s$ and $B_c$. This behaviour confirmed further the almost sole presence of low coercivity minerals (Ti-poor titanomagnetite or titanomaghemite), whose grains progressively unblocked and became paramagnetic. For the four other bricks (SK-8, SK-14, SK-30 and to lesser extent BK-17), $M_{rs}/M_s$ and $B_c$ decreased less or were fairly stable up to about 550°C before to increase sharply due to the Ti-poor titanohematite (Peters and Dekkers, 2003). The increase of the parameters still above 600°C indicated that Ti-poor titanohematite had high unblocking temperatures.

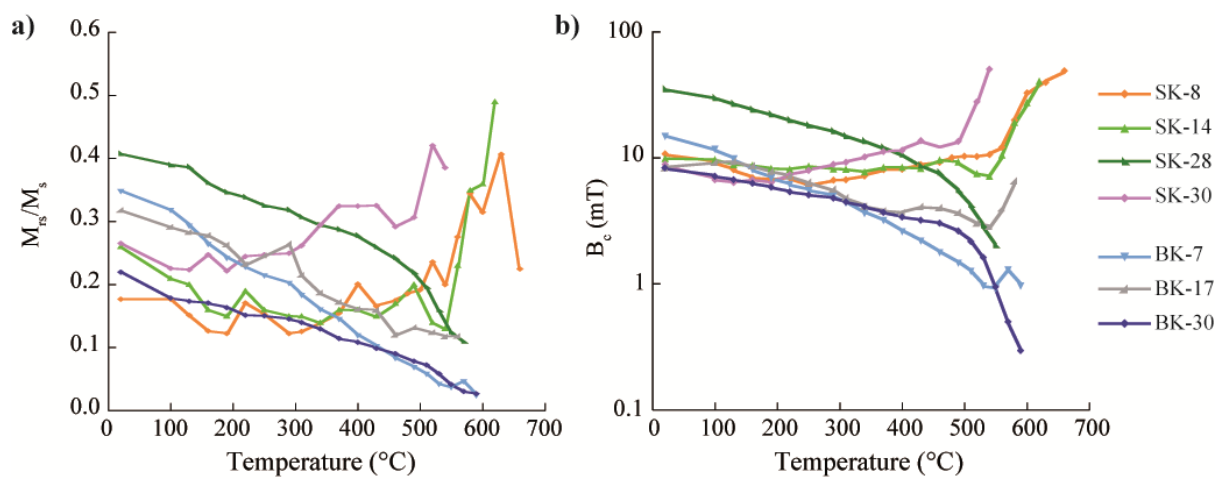

Figure 3: Variation with temperature of (a) the remanence ratio (i.e. the ratio of the saturation remanent magnetization $M_{rs}$ upon the saturation magnetization $M_s$) and (b) the coercive force $B_c$.

To summarize, all rock magnetic experiments showed that the preponderant magnetic carrier was Ti-poor titanomagnetite. Except in six bricks, Ti-poor titanohematite was also present with a higher contribution to the TRM intensity in SK bricks (up to 40%) than in BK bricks. This was likely explained by more oxidizing conditions in the small SK kiln that remained open during baking. No clear relationship appeared between rock magnetic properties and position of the bricks inside each kiln.

4.2 Intensity results

All specimens provided reliable intensity results with single TRM components, linear NRM-TRM diagrams and successful pTRM-checks (Fig. 4, Tabs. 2 and 3). Intensity determinations had high quality factors q between 20 and 101 (Coe *et al.*, 1978). Noteworthy was the slight trend to concave-down NRM-TRM diagrams above 550°C on a few specimens (Fig. 4b).

In the TRM anisotropy protocol, the acceptance limit of the stability check fixed to 5% discarded five specimens (SK-18, SK-26, BK-3, BK-13 and BK-25). For other specimens, anisotropy degrees were between 1.07 and 1.59 with no difference between the two sets of bricks (Fig. S2). As expected for this material, the anisotropy tensor was mostly oblate with a minimal axis $K_{min}$ parallel to the thickness of the brick (e.g. Tema, 2009).

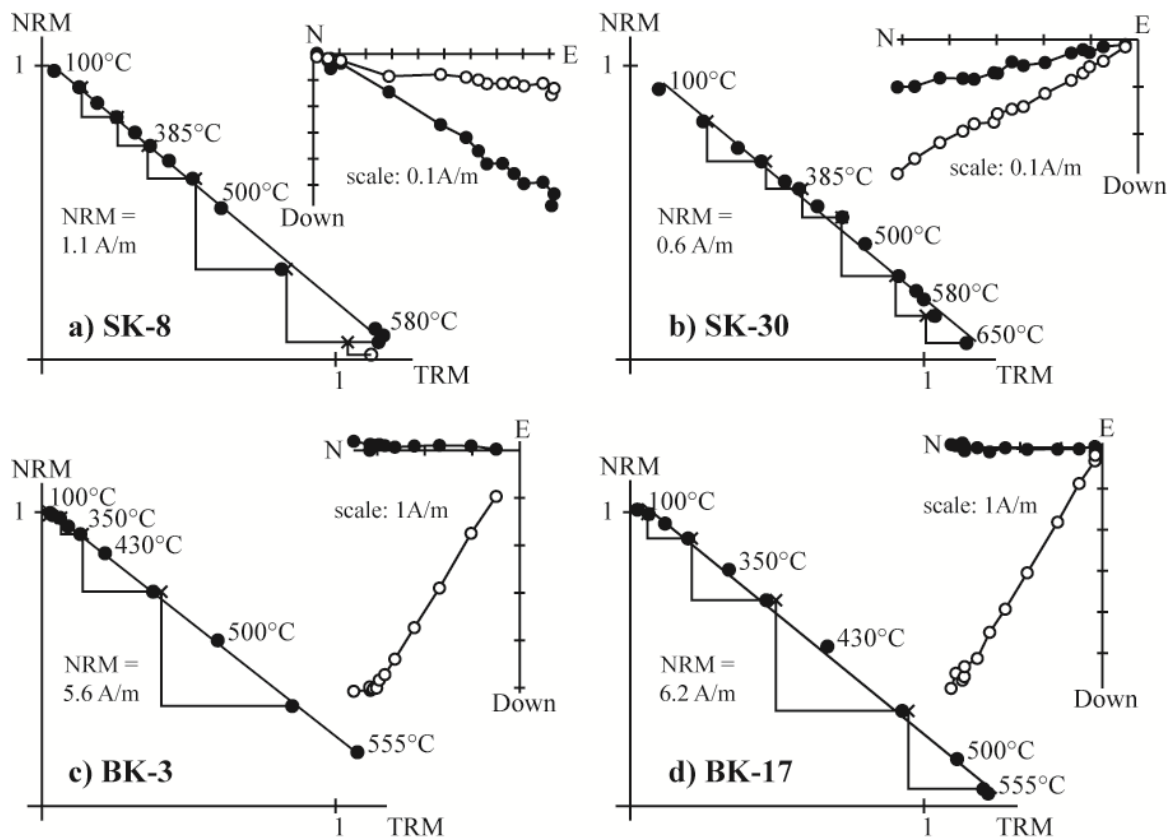

Figure 4: Representative Thellier-Thellier results with NRM-TRM and orthogonal diagrams of SK (a-b) and BK (c-d) bricks. Open (solid) circles on the orthogonal diagrams denote the projection on the vertical (horizontal) plane.

The cooling rate correction was applied when the alteration factor %alt was lower than ±5% and when the absolute value of %corr was higher than the one of %alt (Gomez-Paccard et al., 2006). In the case of SK bricks, %corr$_{0.8°C/min}$ varied between 6.2% and 21.4% and %corr$_{0.4°C/min}$ between 11.3% and 29.2% (Tab. 2). For BK bricks, the values of %corr$_{0.8°C/min}$ were between -2.3% and 7.1% and those of %corr$_{0.4°C/min}$ between -2.1% and 10.1% (Tab. 3). The correction factors of SK bricks were significantly higher than those of BK bricks (Fig. 5).

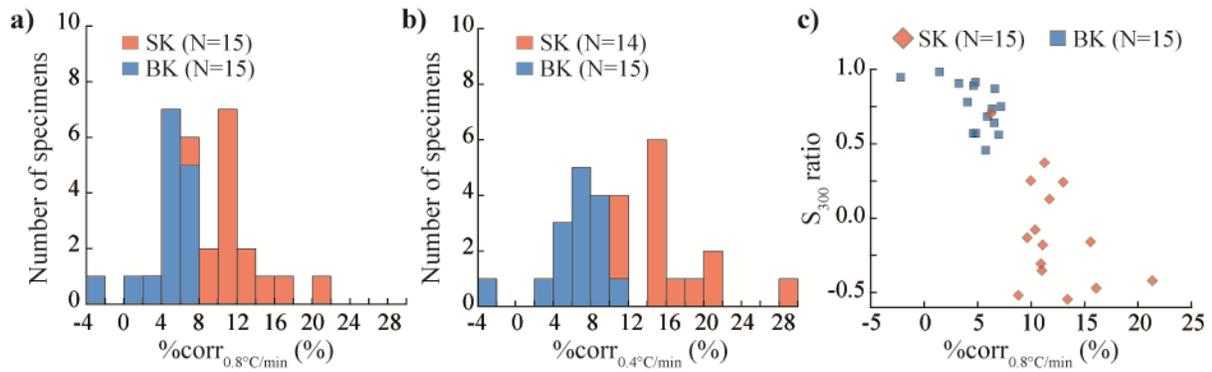

Figure 5: Histograms of cooling rate correction factor %corr for 0.8°C/min (a) and 0.4°C/min (b), cooling rates from 650°C for SK bricks and from 555°C for BK bricks; (c) $S_{300}$ ratio versus cooling rate correction factors (0.8°C/min slow cooling).

The four decreasing cooling rates (0.8, 0.4, 0.2 and 0.1°C/min) performed on BK samples indicated a logarithmic increase of the correction factor with the cooling duration (Fig. 6). This trend was consistent with the theoretical predictions of Halgedahl et al. (1980), which have already been experimentally verified (e.g. Genevey et al., 2008; Muxworthy et al., 2011). The specimen BK-30 with correction factors around -2% did not show a significant variation between 0.8 and 0.4°C/min.

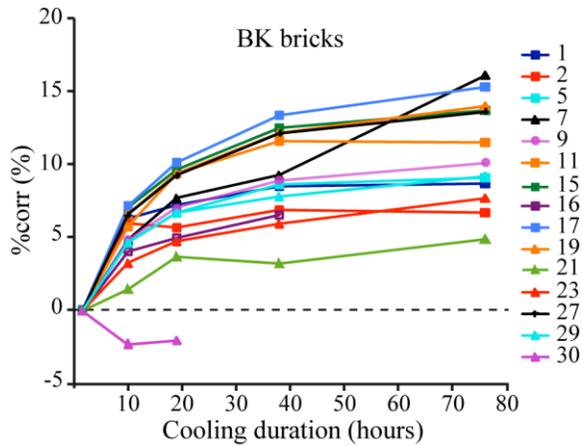

Figure 6: Variation of the cooling rate correction factor with cooling duration in BK bricks. The four durations (10, 19, 38 and 76 hours) corresponded to 0.8°C/min, 0.4°C/min, 0.2°C/min and 0.1°C/min cooling rates respectively.

4.3 Average direction and archaeointensity

The mean direction of the *in situ* oriented BK bricks was calculated using the Fisher (1953) statistic (Fig. 7b). The SK bricks that were put on the side (all bricks except SK-2 and SK-4) allowed the recovery of the inclination (Lanos, 1994). Average inclination was calculated with McFadden and Reid (1982) statistic (Fig. 7a). The anisotropy correction resulted in significant changes of the direction with especially a 5° increase of the inclination in BK bricks (Tab. 2 and 3). The average inclination of SK (58.5±3.0°) and BK (58.2±1.0°) bricks was very close to the expected 58.8° value (Tab. 1). However, the average declination of BK bricks was 4° lower than the IGRF value. This difference was likely due to a local deviation of the magnetic compass during sampling by the remanent magnetization of the kiln walls. Actually the brick BK-30 with the furthest declination from the expected value was the closest to the walls.

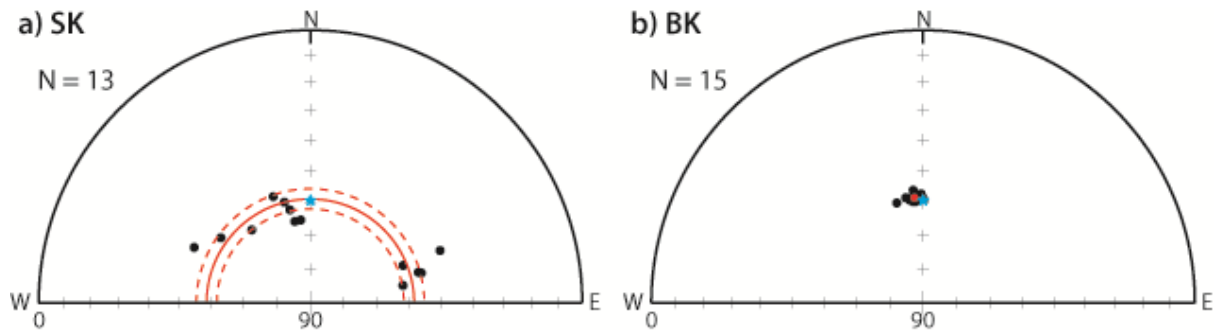

Figure 7: Stereographic plots of directions corrected for TRM anisotropy. For SK bricks (a), the average inclination plotted in red with its 95% error was calculated using McFadden and Reid (1982) statistic. For BK, the average direction in red was calculated using Fisher (1953) statistic. The blue star indicates the direction predicted by IGRF model at the site.

Intensity determinations uncorrected for TRM anisotropy effect provided average values per kiln significantly higher than the expected geomagnetic intensity at the site (21% for SK and 13% for BK kiln). The TRM anisotropy correction decreased the standard deviation of the mean intensity by 68% for SK and by 60% for BK bricks (Tab. 2 and 3). However, mean archaeointensities still overestimated the expected 46.0 µT value by 5-6%.

The choice of the suitable cooling rate depended on the spectra of unblocking temperatures ($T_{ub}$). Dominant $T_{ub}$ in BK bricks ranged from 200-250°C to 550°C (Fig. 2f). Cooling in this temperature interval lasted about 40 hours in the kiln (Fig. 1b) and therefore the choosen cooling rate was 0.2°C/min. The corrected average intensity, 45.1±1.8 µT, was close to 46.0 µT (Tab. 3). It is worth pointing out that others slow cooling rates also provided the expected value when uncertainties were considered.

In the case of SK bricks, the cooling from 700-650°C to 150-100°C took around 12 hours inside the kiln. The corrected average intensity for the 0.8°C/min slow cooling rate was 42.6±2.3 µT, 8% lower than the true value (Tab. 2). SK bricks that underestimated the most the expected intensity presented the highest correction factor (Fig. S3). This suggests that cooling rate correction factors were overestimated.

| Brick | Without corrections | | | | | | | | | | ATRM | | | Cooling rate correction at 650°C | | | | | | CR correction at 580°C | | |
|---|---|---|---|---|---|---|---|---|---|---|---|---|---|---|---|---|---|---|---|---|---|---|
| | | | | | | | | | | | | | | 0.8°C/min (~12 h) | | | 0.4°C/min (~23 h) | | | 0.8°C/min (~11 h) | | |
| | Tmin - Tmax (°C) | n | I (°) | F (µT) | Mad (°) | Dang (°) | f | g | q | ß | Check (%) | $I_a$ (°) | $F_a$ (µT) | %corr | %alt | $F_{a+CR}$ (µT) | %corr | %alt | $F_{a+CR}$ (µT) | $F_{a\ 100-580°C}$ (µT) | %corr | $F_{a+CR\ 100-580°C}$ (µT) |
| SK-2  | 100 - 650 | 14 | -    | 52.1 | 2.9 | 1.3 | 0.84 | 0.91 | 31.8 | 0.024 | -1.3  | -    | 48.7 | 13.0 | -1.7 | 42.4 | 17.6 | -0.4 | 40.1 | 47.7 | 8.8 | 43.5 |
| SK-4  | 100 - 650 | 14 | -    | 52.6 | 4.6 | 1.3 | 0.74 | 0.89 | 41.6 | 0.016 | 0.4   | -    | 47.1 | 11.0 | -1.7 | 41.9 | 15.1 | 0.1  | 40.0 | 46.4 | 4.8 | 44.1 |
| SK-6  | 100 - 650 | 14 | 65.3 | 48.4 | 3.3 | 0.5 | 0.79 | 0.88 | 43.9 | 0.016 | -2.8  | 65.2 | 48.1 | 8.8  | -1.5 | 43.9 | 11.3 | 0.5  | 42.7 | 47.1 | 1.2 | 46.5 |
| SK-8  | 100 - 605 | 13 | 57.7 | 58.9 | 3.0 | 1.3 | 0.90 | 0.86 | 45.0 | 0.017 | -4.5  | 58.9 | 51.0 | 6.2  | -3.7 | 47.8 | 4.1  | -7.9 |      |      |     |      |
| SK-10 | 100 - 650 | 14 | 56.2 | 55.4 | 2.6 | 1.0 | 0.87 | 0.91 | 51.4 | 0.015 | 0.4   | 56.1 | 49.8 | 11.7 | -1.5 | 44.0 | 14.6 | -0.6 | 42.6 | 49.8 | 4.7 | 47.5 |
| SK-12 | 100 - 650 | 14 | 58.4 | 55.2 | 3.8 | 0.5 | 0.81 | 0.91 | 42.3 | 0.017 | 0.5   | 56.6 | 48.4 | 11.1 | -1.4 | 43.0 | 14.6 | -0.4 | 41.3 | 47.8 | 6.0 | 44.9 |
| SK-14 | 100 - 650 | 14 | 50.1 | 53.8 | 3.9 | 1.8 | 0.84 | 0.91 | 41.7 | 0.018 | -1.3  | 47.4 | 48.7 | 11.3 | -1.3 | 43.2 | 14.7 | 0.0  | 41.6 | 47.8 | 4.8 | 45.5 |
| SK-16 | 100 - 650 | 14 | 56.2 | 59.1 | 2.4 | 1.6 | 0.87 | 0.91 | 37.4 | 0.021 | -1.2  | 50.9 | 47.6 | 9.7  | -1.8 | 43.0 | 11.7 | -2.1 | 42.0 | 47.8 | 5.4 | 45.2 |
| SK-18 | 100 - 530 | 10 | 61.0 | 65.5 | 4.2 | 4.3 | 0.61 | 0.82 | 20.1 | 0.025 | -8.0  |      |      |      |      |      |      |      |      |      |     |      |
| SK-20 | 100 - 650 | 14 | 65.5 | 57.3 | 3.6 | 3.1 | 0.83 | 0.89 | 33.7 | 0.022 | -0.5  | 65.0 | 49.0 | 21.4 | -1.0 | 38.5 | 29.2 | 2.5  | 34.7 | 48.8 | 7.2 | 45.3 |
| SK-22 | 100 - 650 | 14 | 61.3 | 52.0 | 4.7 | 3.5 | 0.84 | 0.89 | 62.8 | 0.012 | -1.9  | 55.2 | 46.3 | 16.1 | -0.3 | 38.8 | 21.9 | 2.3  | 36.2 | 46.3 | 3.6 | 44.6 |
| SK-24 | 100 - 650 | 14 | 61.1 | 53.2 | 3.7 | 0.7 | 0.79 | 0.89 | 41.3 | 0.017 | -0.3  | 61.6 | 48.5 | 10.9 | -1.6 | 43.2 | 14.4 | 0.3  | 41.5 | 48.2 | 4.4 | 46.1 |
| SK-26 | 100 - 530 | 10 | 60.6 | 70.5 | 4.1 | 4.3 | 0.76 | 0.81 | 42.6 | 0.014 | -11.2 |      |      |      |      |      |      |      |      |      |     |      |
| SK-28 | 100 - 650 | 14 | 63.0 | 57.4 | 2.5 | 1.3 | 0.84 | 0.91 | 75.4 | 0.010 | -2.0  | 61.8 | 48.1 | 10.0 | -1.8 | 43.3 | 11.0 | -1.5 | 42.8 | 48.6 | 2.7 | 47.3 |
| SK-30 | 100 - 650 | 14 | 65.2 | 59.2 | 2.7 | 1.1 | 0.84 | 0.91 | 31.4 | 0.024 | -1.5  | 59.9 | 48.7 | 15.6 | -0.6 | 41.1 | 20.5 | 0.4  | 38.7 | 47.4 | 7.7 | 43.7 |
| SK-32 | 100 - 650 | 14 | 60.0 | 54.4 | 3.1 | 1.2 | 0.77 | 0.88 | 34.3 | 0.020 | 0.0   | 56.1 | 45.7 | 13.4 | -0.6 | 39.6 | 18.9 | 0.9  | 37.1 | 44.2 | 7.4 | 41.0 |
| SK-34 | 100 - 650 | 14 | 64.9 | 54.9 | 2.6 | 2.0 | 0.78 | 0.91 | 46.1 | 0.016 | -2.3  | 61.7 | 48.9 | 10.4 | -0.8 | 43.8 | 14.4 | 0.8  | 41.9 | 48.5 | 6.4 | 45.4 |
| Average inclination (McFadden & Reid, 1982) | $I_{mean}$ = 60.5°  N = 15  k = 185  $\alpha_{95}$ = 2.3° | | | | | | | | | | $I_{mean}$ = 58.2°  N = 13  k = 120  $\alpha_{95}$ = 3.8° | | | | | | | | | | | |
| Average weigthed intensity $F_{mean}$ | $F_{mean}$ = 56.4 ± 5.3 µT   N = 17 | | | | | | | | | | $F_{mean}$ = 48.3 ± 1.3 µT  N = 15 | | | $F_{mean}$ = 42.6 ± 2.3 µT  N = 15 | | | $F_{mean}$ = 40.5 ± 2.6 µT  N = 14 | | | $F_{mean}$ = 45.0 ± 1.7 µT   N = 14 | | |
| IEF (%) | +23% | | | | | | | | | | +5% | | | -8% | | | -12% | | | -2% | | |

Table 2: Intensity results of bricks from SK small kiln. The cooling rate closer to the initial one is highlighted in grey.

Above, from left to right: Name of the brick; Temperature interval used to determine intensity; Number of temperature steps within $T_{min}$-$T_{max}$; Inclination and Intensity without corrections; Maximum angular deviation MAD; Deviation angle DANG; NRM fraction f; gap factor g; quality factor q (Coe *et al.*, 1978); Ratio of the standard error of the slope to the absolute value of the slope ß; Stability check of the anisotropy of TRM (ATRM) correction; Inclination and intensity corrected for ATRM; correction factor %corr, alteration factor %alt and intensity corrected for ATRM and cooling rate effects for the two tested slow cooling rates; Intensity determined up to 580°C with anisotropy correction alone, cooling rate correction factor below 580°C and intensity up to 580°C with both corrections.

Below, rows from top: Average inclination calculated using MacFadden & Reid (1982) statistics from N bricks with precision parameter k and semi-angle of confidence at 95 per cent $\alpha_{95}$; Average intensity with standard deviation from N bricks; the intensity error fraction, IEF, is the deviation in percentage of the average intensity to the true 46.0 µT value.

| | Without corrections | | | | | | | | | | | ATRM | | | | Cooling rate correction (at 555°C) | | | | | | | | | | | |
|---|---|---|---|---|---|---|---|---|---|---|---|---|---|---|---|---|---|---|---|---|---|---|---|---|---|---|---|
| | | | | | | | | | | | | | | | | 0.8°C/min (~10 h) | | | 0.4°C/min (~19 h) | | | 0.2°C/min (~38 h) | | | 0.1°C/min (~76 h) | | |
| Brick | $T_{min}$-$T_{max}$ (°C) | n | D (°) | I (°) | F (µT) | Mad (°) | Dang (°) | f | g | q | ß | Check (%) | $D_a$(°) | $I_a$ (°) | $F_a$ (µT) | %corr | %alt | $F_{a+CR}$ (µT) | %corr | %alt | $F_{a+CR}$ (µT) | %corr | %alt | $F_{a+CR}$ (µT) | %corr | %alt | $F_{a+CR}$ (µT) |
| BK-1 | 100 - 555 | 11 | 350.6 | 56.6 | 48.1 | 1.5 | 1.0 | 0.92 | 0.77 | 40.4 | 0.018 | -0.6 | 353.9 | 58.3 | 49.7 | 6.3 | -0.2 | 46.6 | 7.2 | -0.6 | 46.1 | 8.5 | -0.6 | 45.5 | 8.7 | -2.2 | 45.4 |
| BK-2 | 100 - 555 | 11 | 348.4 | 56.2 | 49.9 | 1.6 | 0.6 | 0.94 | 0.85 | 40.5 | 0.020 | -2.0 | 356.0 | 59.3 | 49.5 | 5.9 | 0.0 | 46.6 | 5.7 | -1.5 | 46.7 | 6.9 | -2.6 | 46.1 | 6.7 | -5.0 | 46.2 |
| BK-3 | 100 - 555 | 11 | 354.9 | 53.8 | 50.8 | 2.0 | 1.4 | 0.80 | 0.82 | 101.4 | 0.007 | -6.8 | | | | | | | | | | | | | | | |
| BK-5 | 100 - 555 | 11 | 360.0 | 53.4 | 48.8 | 2.5 | 0.2 | 0.89 | 0.88 | 56.3 | 0.014 | -0.2 | 355.0 | 55.8 | 49.3 | 4.6 | 0.1 | 47.0 | 6.7 | 0.5 | 46.0 | 8.6 | 0.2 | 45.1 | 9.1 | 0.4 | 44.8 |
| BK-7 | 100 - 555 | 11 | 356.3 | 53.6 | 51.5 | 2.7 | 1.2 | 0.90 | 0.90 | 56.9 | 0.014 | -1.7 | 355.2 | 57.8 | 48.9 | 4.8 | 0.0 | 46.6 | 7.7 | 0.6 | 45.1 | 9.2 | -1.1 | 44.4 | 16.1 | -0.6 | 41.0 |
| BK-9 | 100 - 555 | 11 | 358.1 | 51.0 | 55.1 | 1.5 | 1.0 | 0.95 | 0.84 | 57.8 | 0.014 | 0.9 | 360.0 | 58.4 | 48.5 | 4.8 | 0.1 | 46.2 | 7.0 | 0.5 | 45.1 | 8.9 | 0.2 | 44.2 | 10.1 | 0.3 | 43.6 |
| BK-11 | 100 - 555 | 11 | 355.5 | 52.4 | 52.0 | 2.7 | 0.3 | 0.86 | 0.88 | 32.9 | 0.023 | -0.5 | 358.1 | 57.6 | 47.3 | 5.7 | 0.2 | 44.6 | 9.3 | 1.4 | 42.9 | 11.6 | 0.6 | 41.8 | 11.5 | 1.2 | 41.9 |
| BK-13 | 100 - 530 | 10 | 356.3 | 54.4 | 49.5 | 2.3 | 2.7 | 0.70 | 0.77 | 46.7 | 0.012 | -15.9 | | | | | | | | | | | | | | | |
| BK-15 | 100 - 555 | 11 | 359.0 | 53.1 | 51.6 | 1.9 | 1.0 | 0.85 | 0.88 | 38.1 | 0.020 | -0.4 | 353.9 | 57.9 | 48.5 | 7.0 | 0.5 | 45.1 | 9.6 | 1.2 | 43.8 | 12.5 | 1.1 | 42.4 | 13.7 | 2.0 | 41.9 |
| BK-16 | 100 - 530 | 10 | 353.4 | 52.5 | 51.1 | 3.0 | 1.0 | 0.74 | 0.88 | 82.8 | 0.008 | -3.8 | 352.7 | 57.0 | 48.1 | 4.0 | 0.1 | 46.2 | 4.9 | -0.3 | 45.7 | 6.5 | -2.8 | 45.0 | 7.2 | -5.6 | |
| BK-17 | 100 - 555 | 11 | 5.8 | 53.4 | 52.6 | 1.8 | 0.6 | 0.91 | 0.86 | 45.5 | 0.017 | 0.1 | 0.4 | 58.8 | 49.4 | 7.1 | 0.1 | 45.9 | 10.1 | 0.7 | 44.4 | 13.3 | 0.8 | 42.8 | 15.3 | 1.7 | 41.8 |
| BK-19 | 100 - 555 | 11 | 356.6 | 52.1 | 52.0 | 1.1 | 1.1 | 0.93 | 0.85 | 63.4 | 0.012 | 0.2 | 356.2 | 58.2 | 51.4 | 6.6 | -0.4 | 48.0 | 9.3 | 0.3 | 46.6 | 12.1 | 0.5 | 45.2 | 14.0 | 1.0 | 44.2 |
| BK-21 | 100 - 555 | 11 | 2.9 | 44.7 | 59.6 | 1.7 | 1.3 | 0.87 | 0.85 | 42.1 | 0.017 | -2.8 | 358.2 | 57.0 | 49.7 | 1.4 | -1.2 | 49.0 | 3.7 | -1.3 | 47.9 | 3.2 | -1.7 | 48.1 | 4.9 | -3.9 | 47.3 |
| BK-23 | 100 - 555 | 11 | 358.2 | 54.2 | 53.5 | 1.8 | 0.9 | 0.97 | 0.84 | 90.2 | 0.009 | -0.1 | 354.5 | 59.3 | 48.3 | 3.2 | -0.3 | 46.8 | 4.7 | -0.2 | 46.0 | 5.9 | 0.2 | 45.5 | 7.7 | -0.2 | 44.6 |
| BK-25 | 100 - 555 | 10 | 0.1 | 51.2 | 48.4 | 2.9 | 0.8 | 0.56 | 0.83 | 35.8 | 0.013 | -6.1 | | | | | | | | | | | | | | | |
| BK-27 | 100 - 555 | 11 | 2.0 | 57.5 | 50.1 | 1.5 | 1.0 | 0.90 | 0.86 | 40.2 | 0.019 | 0.2 | 352.7 | 58.7 | 49.4 | 6.5 | -0.1 | 46.2 | 9.2 | 0.7 | 44.9 | 12.1 | 0.7 | 43.4 | 13.6 | 1.7 | 42.7 |
| BK-29 | 100 - 555 | 11 | 357.0 | 49.4 | 57.0 | 1.4 | 0.2 | 0.96 | 0.83 | 74.7 | 0.011 | -0.4 | 353.5 | 58.9 | 49.7 | 4.6 | -1.5 | 47.4 | 6.7 | -0.6 | 46.4 | 7.8 | -1.4 | 45.8 | 9.2 | -1.8 | 45.1 |
| BK-30 | 100 - 555 | 11 | 341.8 | 51.9 | 52.1 | 4.6 | 0.2 | 0.60 | 0.82 | 66.8 | 0.007 | -4.4 | 345.4 | 58.8 | 47.2 | -2.3 | -0.7 | 48.3 | -2.1 | -0.2 | 48.2 | -2.0 | -2.9 | | -1.7 | -5.1 | |
| Average direction (Fisher, 1953) | $D_{mean}$ = 356.6° $I_{mean}$ = 53.3° N = 18 k = 338 $\alpha_{95}$ = 1.9° | | | | | | | | | | | $D_{mean}$ = 355.1° $I_{mean}$ = 58.2° N = 15 k = 1442 $\alpha_{95}$ = 1.0° | | | | | | | | | | | | | | | |
| Average weighted intensity $F_{mean}$ | $F_{mean}$ = 52.1 ± 3.0 µT  N = 18 | | | | | | | | | | | $F_{mean}$ = 48.9 ± 1.1 µT  N = 15 | | | | $F_{mean}$ = 46.8 ± 1.2 µT N = 15 | | | $F_{mean}$ = 45.9 ± 1.4 µT N = 15 | | | $F_{mean}$ = 45.1 ± 1.8 µT N = 15 | | | $F_{mean}$ = 44.0 ± 1.9 µT N = 13 | | |
| IEF (%) | +13% | | | | | | | | | | | +6% | | | | +2% | | | 0% | | | -2% | | | -4% | | |

Table 3: Intensity results of bricks from BK big kiln. The cooling rate closer to the initial one is highlighted in grey.

Above, from left to right: Name of the brick; Temperature interval used to determine intensity; Number of temperature steps within $T_{min}$-$T_{max}$; Declination, inclination and Intensity without corrections; see Table 2 for MAD, DANG, f, g, q and ß; Stability check of the anisotropy of TRM (ATRM) correction; Declination, inclination and intensity corrected for ATRM; correction factor %corr, alteration factor %alt and intensity corrected for ATRM and cooling rate for each tested slow cooling rates.

Below, rows from top: Average declination and inclination calculated using Fisher (1953) statistics from N bricks with precision parameter k and semi-angle of confidence at 95 per cent $\alpha_{95}$; Average intensity with standard deviation from N bricks; the intensity error fraction, IEF, is the deviation in percentage of the average intensity to the true 46.0 µT value.

5. Discussion

5.1 Cooling rate effect and rock magnetic properties

The cooling rate effect upon the acquired TRM intensity is clearly related to the ferromagnetic mineralogy. Here, TRM intensity increased with slower coolings in all cases, except for the brick with titanomaghemite (BK-30) (Tabs. 2 and 3). According to high field experiments, Ti-poor titanomagnetites were present in all samples. TRM fraction carried by Ti-poor titanohematite grains was large in SK bricks and not significant in BK bricks. Our results showed that $S_{300}$ ratio was inversely proportional to the cooling rate correction factors (Fig. 5c). Samples with $S_{300}$ close to 1 have %corr$_{0.8°C/min}$ around 0%, those with $S_{300}$ between 0.5 and 0.95 had generally %corr$_{0.8°C/min}$ between 5 and 10% and those with $S_{300}$ below 0.5 have %corr$_{0.8°C/min}$ higher than 10%. This trend suggests that the TRM intensity acquired by high coercivity grains (i.e. Ti-poor titanohematites) was more dependent to the cooling rate than the one acquired by Ti-poor magnetites.

The higher dependency of Ti-poor hematites is also proved with the variation of the cooling correction factor with temperature ranges (Fig. 8a). On SK bricks, the %corr$_{0.8°C/min}$ factors for the intervals 385-20°C, 500-20°C and 580-20°C were small with respective average values of 4.3 ± 0.9 %, 2.7 ± 1.4 % and 5.4 ± 2.1 %. The average correction factor from 650°C to 20°C was more than twice higher, 12.5 ± 3.3 %, indicating that this increase was due to the TRM acquired by Ti-poor titanohematites.

In order to further ascertain the role of this phase, we performed an additional experiment from 650°C to room temperature with a slow cooling rate (0.8°C/min) only applied from 650°C to 580°C. While only 10-26% of the NRM demagnetized between 580°C and 650°C, the correction factor in this temperature interval corresponded to 39%-75% of the correction factor between 650 and 20°C (Fig. 8c). Ti-poor titanohematite grains contributed three times more to the cooling rate effect than to the NRM intensity.

In the case of BK bricks, mean %corr$_{0.8°C/min}$ factors from 350°C and from 430°C to room temperature were similar, 3.3 ± 1.2 % and 3.1 ± 1.0 % respectively (Fig. 8b). Trends varied between specimens for higher temperatures, factors being stable or increasing from 430°C to 500°C steps and then increasing or decreasing at 550°C step for most bricks. Average correction factors from 500°C and 550°C to room temperature were respectively 5.3±2.2% and 4.7±2.5%. The variability of behaviours can be explained by slight evolutions (a few per cent, see Tab. 3) during the 0.4°C/min, 0.2°C/min and 0.1°C/min slow coolings from 550°C that were carried out before the 0.8°C/min slow coolings from 350°C, 430°C and 500°C. Taking into account this possibility, we cannot conclude that the cooling rate effect increased close to the Curie temperature, as observed by Papusoi (1972a) and Yu (2011).

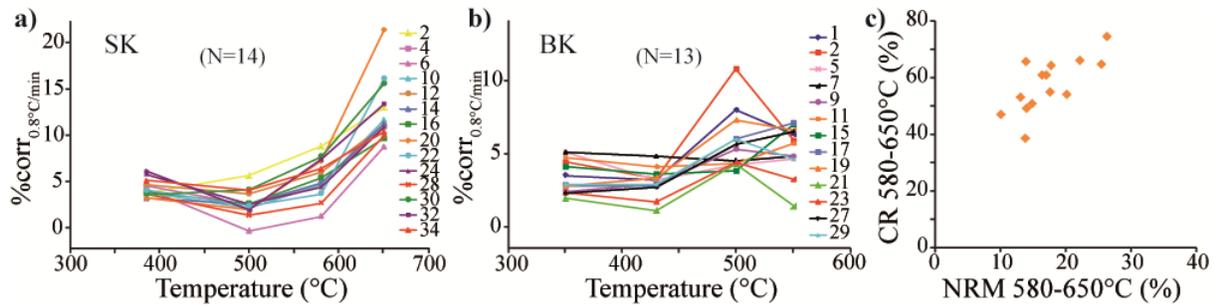

Figure 8: Variation of the cooling rate correction factors (0.8°C/min slow cooling) with temperature in SK (a) and BK (b) bricks. (c) For SK bricks, fraction of NRM demagnetized in this temperature interval versus percentage of the correction factor in the temperature interval 580-650°C.

Theoretical studies and experiments on synthetic grains demonstrated a decrease of the cooling rate effect with increasing grain size, from SD to MD grains, both for magnetite and hematite (Halgedahl *et al.*, 1980; McClelland, 1984; Papusoi, 1974). In Sallèles-d'Aude bricks, this relationship would imply that Ti-poor titanohematite grains would be finer than those of Ti-poor titanomagnetite. Verifying this hypothesis was difficult, because mixing of ferromagnetic minerals in the bricks prevented the determination of the grain size using hysteresis parameters. However, linear NRM-TRM diagrams discarded that multidomain grains carried a significant part of the TRM (e.g. Levi, 1977).

Independently of the grain size, one may wonder if the cooling rate effect could be intrinsically higher for Ti-poor titanohematites than for Ti-poor titanomagnetites. A slower cooling rate lowers the blocking temperature, which results in a higher spontaneous magnetization $M_s$ when magnetic moment of grains blocked and therefore in a higher TRM intensity (Dodson and

McClelland, 1980). The increase of $M_s$ below Curie temperature being sharper for Ti-poor titanohematites than for Ti-poor titanomagnetites (e.g. Dunlop & Özdemir, 1997), a decrease of blocking temperatures may lead to a larger increase of TRM intensity acquired by Ti-poor titanohematites. Testing this hypothesis, using for example experiments on synthetic minerals, might open the way to a possible mineralogy-dependent cooling rate correction.

Finally, it is also possible that a linear decrease of the temperature (imposed by technical constraints) during laboratory experiments was not the best choice to mimic the initial cooling of the bricks in archaeological kilns, especially in the case of baked clays containing a significant amount of ferromagnetic minerals with high blocking temperatures. In Figure 1, the monitored cooling clearly follows a rather logarithmic curve than a linear one.

5.2 Accuracy of the cooling rate correction

In some samples (e.g. SK-30, Fig. 4b), the increase of the cooling rate effect with temperature resulted in slightly concave-down NRM-TRM plots above 550°C. In SK bricks, the higher contribution of Ti-poor titanohematite grains to the cooling rate correction than to the TRM intensity (Fig. 8c) explained the overestimate of the correction factor and the underestimate by 8% of the archaeointensity calculated over all the temperature range. Actually, the archaeointensities determined between 100°C and 580°C and corrected for the cooling rate only from 580°C, gave an average value of 45.0 ± 1.7 µT, very close to the expected one (Tab. 2).

The example of SK bricks shows that archaeointensity data have to be treated more cautiously when Ti-poor titanohematites carry a significant part of the TRM. In this case, performing the cooling rate correction before and after the Curie temperature of magnetite allows checking the accuracy of the archaeointensity results. Using the Triaxe protocol is also useful, because this method is not sensitive to the cooling rate effect (Le Goff and Gallet, 2004). Fortunately, a high contribution of Ti-poor titanohematites to TRM rarely occurs in archaeological baked clays..

Generally, the true past cooling rate of archaeological kilns is unknown. Experimental archaeology can only provide a rough estimation based on the kiln sizes. However, our intensity results on BK bricks showed that corrected average intensities after slow coolings over ~10, ~19, ~38 and ~76 hours were close to the expected value when the uncertainties are considered (Tab. 3). As all corrected average values were more accurate than the one after only TRM anisotropy correction, applying the cooling rate correction was in any case the best approach.

7. Conclusions

We studied two sets of bricks, modelled in the same clay and baked in two modern kilns (BK and SK). The ambient geomagnetic field, maximal reached temperatures and cooling durations were recorded at the site. Rock magnetic experiments demonstrated that TRM is carried by Ti-poor titanomagnetites in both sets of bricks. Most of the bricks from the smallest kiln, SK, contained also a significant amount of high coercivity ferromagnetic grains, likely Ti-poor titanohematites.

All bricks showed thermal stability of ferromagnetic mineralogy and univectorial TRM. Archaeointensities were determined using the Thellier-Thellier protocol with TRM anisotropy and cooling rate corrections. The knowledge of experimental heating and cooling conditions of the bricks allowed us to investigate the cooling rate effect on intensity results. For this, we applied several slow cooling rates, 0.8, 0.4, 0.2 and 0.1°C/min successively, during the Thellier-Thellier experiments.

Main results are:

- After correction for TRM anisotropy, the average directions of the bricks recovered the expected geomagnetic field direction but the average intensities overestimated the expected value by 5-6%.

- Cooling of the BK kiln in the interval of unblocking temperatures of bricks lasted about 40 hours. The expected intensity (46.0 µT) was obtained on BK bricks after cooling rate correction using four cooling durations between 10 and 76 hours. This suggests that a precise knowledge of the archaeological cooling duration in a kiln is not essential for an efficient correction of the cooling rate effect on TRM intensity.

- The strongest dependency of the TRM intensity on cooling rate was observed for SK kiln bricks (containing Ti-poor titanohematites). This discrepancy might be due to the chemical nature of the grains rather than to their size. It is known that the increase of $M_s$ below Curie temperature is sharper for Ti-poor titanohematites than for Ti-poor titanomagnetites. As a consequence, the decrease of blocking temperatures induced by a slower cooling rate would lead to a larger increase of TRM intensity acquired by Ti-poor titanohematites.

- SK kiln cooled from 700-650°C to 150-100°C in 12 hours. Applying the corresponding 0.8°C/min slow cooling in this temperature interval led to an overestimate of the cooling rate correction and then to an underestimate of the geomagnetic field intensity by 8%. The expected intensity was only recovered when intensity was determined and cooling rate correction applied only between 580°C and 100°C.

- The example of SK bricks shows that archaeointensity data have to be treated more cautiously when Ti-poor titanohematites carry a significant part of the TRM. In this case, performing the cooling rate correction before and after the Curie temperature of magnetite is useful to check the accuracy of the archaeointensity results. Another possibility would be to avoid a linear decrease of the temperature with time over all the temperature range. A progressive decrease of the cooling rate from high (>580°C) to room remperatures would better mimic the cooling of archaeological kilns. Setting up such experimental monitoring is unfortunately not straightforward in the laboratory.

Acknowledgements

Funding was providing to GH by Campus France PRESTIGE program (PRESTIGE-2017-1-0002). GH is also supported by ANR-CONACYT SVPIntMex project ANR-15-CE31-0011-01 (coord. Mireille Perrin, CEREGE and Luis Alva-Valdivia, UNAM). We thank Yongjae Yu and Wilbor Poletti for their detailed reviews.